\documentclass[prd,twocolumn,notitlepage,preprintnumbers,amsmath,amssymb,nofootinbib,10pt,superscriptaddress]{revtex4-2}

\usepackage{dcolumn}
\usepackage{bm}
\usepackage{xcolor}
\usepackage[utf8]{inputenc}
\usepackage[spanish,english]{babel}
\usepackage{amsmath,amssymb,amsfonts,latexsym,cancel}
\usepackage{graphicx}
\usepackage{color}
\usepackage{soul}
\usepackage[normalem]{ulem}
\usepackage{hyperref}
\usepackage{amsmath}
\usepackage{slashed}
\usepackage{braket}
\usepackage{amssymb}
\usepackage{amsmath}
\graphicspath{ {./images/} }

\newcommand{\be}{\begin{equation}}
\newcommand{\ee}{\end{equation}}
\newcommand{\bea}{\begin{eqnarray}}
\newcommand{\eea}{\end{eqnarray}}

\newcommand{\roverline}[1]{\mathpalette\doroverline{#1}}
\newcommand{\doroverline}[2]{\overline{#1#2}}

\newcommand{\langleRE}{\langle{:}\,}
\newcommand{\rangleRE}{\,{:}\rangle}

\begin{document}

\title{Ultraviolet-regularized power spectrum  without infrared distortions  \\ 
in cosmological spacetimes} 

\author{Antonio Ferreiro}\email{antonio.ferreiro@dcu.ie}
\affiliation{Centre for Astrophysics and Relativity, School of Mathematical Sciences,
Dublin City University, Glasnevin, Dublin 9, Rep. of Ireland.}
\affiliation{Departamento de F\'isica Te\'orica, Universidad de Valencia and Instituto de F\'isica Corpuscular –
CSIC/Universidad de Valencia, Burjassot–46100, Spain.}

\author{Francisco Torrenti}\email{f.torrenti@uv.es}
\affiliation{Departamento de F\'isica Te\'orica, Universidad de Valencia and Instituto de F\'isica Corpuscular –
CSIC/Universidad de Valencia, Burjassot–46100, Spain.}
\affiliation{Department of Physics, University of Basel, Klingelbergstr. 82, CH-4056 Basel, Switzerland.}

\begin{abstract}
We reexamine the regularization of the two-point function of a scalar field in a Friedmann-Lemaitre-Robertson-Walker (FLRW) spacetime. Adiabatic regularization provides a set of subtraction terms in momentum space that successfully remove its ultraviolet divergences at coincident points, but can significantly distort the power spectrum at infrared scales, especially for light fields. In this work we propose, by using the intrinsic ambiguities of the renormalization program, a new set of subtraction terms that minimize the distortions for scales $k \lesssim M$, with $M$ an arbitrary mass scale. Our method is consistent with local covariance and equivalent to general regularization methods in curved spacetime. We apply our results to the regularization of the power spectrum in de Sitter space: while the adiabatic scheme yields exactly $\Delta_{\phi}^{\rm (reg)} = 0$ for a massless field, our proposed prescription recovers the standard scale-invariant result $\Delta_{\phi}^{\rm (reg)} \simeq H^2 /(4\pi^2)$ at super-horizon scales.
\end{abstract}

\date{\today}
\maketitle

\section{Introduction}

Quantum field theory in curved spacetime provides many important insights into the interaction of matter and gravity  \cite{Birrell:1982ix,Parker:2009uva,Wald:1995yp}. An unavoidable consequence of the theory is the amplification of field and metric fluctuations during inflation \cite{Guth:1980zm,Linde:1981mu,Albrecht:1982wi,Starobinsky:1980te}, which later become the seeds for the structure formation in the cosmos \cite{Sasaki:1986hm,Mukhanov:1988jd}. These fluctuations get imprinted in the anisotropies of the Cosmic Microwave Background (CMB), which provide an excellent observational window to the physics of the early universe \cite{Akrami:2018jri,BICEP:2021xfz}.

In this work we consider a free scalar field $\phi$ propagating in a curved spacetime. In the simplest model realizations of inflation, the accelerated expansion of the universe is indeed generated by a scalar field. For (quasi-free) Gaussian states, the information about the state is encoded in the two-point correlation function
\bea
G(x,x'):=\langle \phi(x) \phi(x')\rangle \ , \label{unrG} 
\eea
which acts as a building block for the construction of other relevant quantities such as the stress-energy tensor in the coincident limit $x'\to x$. For instance, the trace of such tensor can be written as
\be
\langle T_a^a(x)\rangle=\left(3\left(\xi-\frac16\right)\Box+m^2\right)\langle \phi^2(x)\rangle \ , 
\ee
where we have used the signature convention $(+,-,-,-)$ (the same as in \cite{Birrell:1982ix,Parker:2009uva}). The quantity $\langle \phi^2(x)\rangle$ contains both quadratic and logarithmic ultraviolet (UV) divergences, which must be removed in order to obtain a physical finite quantity. Several regularization and/or renormalization methods have been developed for this purpose, see e.g.~\cite{Birrell:1982ix,Wald:1995yp,Parker:2009uva} for reviews on the subject. We can use for instance the \textit{point-splitting} regularization method \cite{Christensen:1976vb} to isolate the singular part of the correlation function $G_s (x,x')$ and remove it in a local and covariant manner, before taking the coincident point limit. The regularized two-point function can then be defined as
\be
\langleRE \phi^2(x) \rangleRE=\lim_{x'\to x}\left(G(x,x')-G_s(x,x')\right) \ . \label{renG}
\ee
Nevertheless, other regularization prescriptions can yield different expressions for $\langleRE \phi^2(x) \rangleRE$. A fundamental result of renormalization theory in curved spacetime is that two regularized two-point functions $\langleRE \phi^2(x) \rangleRE$ and $\langleRE\overline{\phi^2(x)} \rangleRE$, regularized with two different methods, are unique up to two arbitrary geometrical contributions of the form 
\be
\langleRE\overline{\phi^2(x)} \rangleRE - \langleRE \phi^2(x) \rangleRE = \alpha m^2+ \beta R \ , \label{difreg}
\ee
where $\alpha$ and $\beta$ are two dimensionless parameters, $R$ is the Ricci scalar, and $m^2$ is the mass squared of the field (see for example \cite{Hollands:2001nf}). This result comes from imposing the minimum requirements of covariance, locality and analiticity in the limit $m^2\to0$.

In the case of a Friedmann-Lemaitre-Robertson-Walker (FLRW) metric, $ds^2=a^2(\tau)(d\tau^2-d\vec{x}^2)$, the symmetries of spacetime allow us to express \eqref{unrG} in terms of fields modes as \cite{Lueders:1990np}
\bea
G(\vec{x},\tau,\vec{x}',\tau')=\frac{1}{(2\pi)^3}\int d^3\vec{k} \, \frac{\chi_k(\tau)}{a(\tau)}\frac{\chi^*_k(\tau')}{a(\tau')}e^{i\vec{k}(\vec{x}-\vec{x}')}.\label{Gcosmo}
\eea
An important quantity that (partially) encodes the information about the two point correlation function is its \textit{power spectrum $P_{k}(\tau)$}, defined in terms of its Fourier transform at equal times $\tau = \tau '$ as
\bea
\mathcal{G}(\tau,k,\tau,k') = (2\pi)^3 \delta^2(k-k')\mathcal{P}_{k}(\tau) ,
\eea
where $k=|\vec{k}|$. Again, \eqref{Gcosmo} is divergent in the limit $\vec{x} \rightarrow \vec{x'}$. This formulation in terms of momentum space allows to envisage a computationally convenient regularization method, consisting in adding an appropriate set of subtraction terms $\mathcal{Q}_{\vec{k}}(\tau)$ inside the integral as follows,
\bea
\langleRE \phi^2(\tau) \rangleRE=\frac{1}{(2\pi)^3}\int d^3\vec{k} \left(\mathcal{P}_{k}(\tau)-\mathcal{Q}_{k}(\tau)\right), \label{eq:integral}
\eea
such that the UV divergences get cancelled and the integral is finite. We define the \textit{regularized power spectrum} as $\mathcal{P}_{k}^{\rm (reg)} \equiv \, \mathcal{P}_{k}-\mathcal{Q}_{k}$.

A well-established regularization method in FLRW spacetimes is \textit{adiabatic regularization}. It is based on an adiabatic expansion of the field modes \cite{Parker:1968mv,Parker:1969au,Parker:1971pt}, which captures their ultraviolet behavior and allows to isolate their divergent contributions to the quantity one wishes to regularize. When applied to the two-point function, the method yields a set of subtraction terms $\mathcal{Q}_{k}(\tau)$ that renders the integral (\ref{eq:integral}) finite (the specific form of $\mathcal{Q}_{k}(\tau)$ is given in Eq.~(\ref{Hk1}) below). The method can also be applied to the regularization of the stress-energy tensor \cite{Parker:1974qw,Fulling:1974zr,Fulling:1974pu}, as well as to other field species such as fermions \cite{Landete:2013axa, Landete:2013lpa, delRio:2014cha, BarberoG:2018oqi}. The adiabatic scheme can also be extended to include interactions of quantum fields to classical time-dependent homogeneous backgrounds, see e.g.~\cite{Anderson:2008dg,Anderson:2015yda, delRio:2017iib} for preheating-like scenarios \cite{Kofman:1994rk,Kofman:1997yn,Greene:1997fu}, as well as \cite{Ferreiro:2018qdi,Ferreiro:2018qzr,Beltran-Palau:2020hdr} for the case of interactions to classical electric fields. The adiabatic method has also been applied to the construction of preferred vacuum states in cosmological spacetimes \cite{Agullo:2014ica, Ferreiro:2022hik}. A fundamental result of the adiabatic scheme is that, even though it assumes a pre-existing FLRW spacetime, regularization techniques in general curved backgrounds converge to equivalent subtraction terms when restricted to a FLRW spacetime \cite{delRio:2014bpa, Beltran-Palau:2021xrk}. This ensures that all obtained observables are constructed in a local covariant way.

Nevertheless, although the subtraction terms obtained with the adiabatic method successfully remove the UV divergences of the two-point function, they can significantly distort the infrared part of the power spectrum, especially for light fields. This effect is clearly seen in the case of a scalar field in de Sitter spacetime: the regularized power spectrum of a light field ($m\ll H$) gets significantly suppressed at scales  $m \ll ka \, (\lesssim H)$, while the one of a massless field \textit{vanishes at all scales} \cite{Parker:2007ni,Agullo:2008ka, Agullo:2009vq,Agullo:2009zi}. These results dramatically change the standard observable predictions of slow-roll inflation, and sparked a fascinating debate on how to correctly apply the regularization program to the inflationary perturbations \cite{Finelli:2007fr,Durrer:2009ii,Urakawa:2009xaa,Marozzi:2011da,Agullo:2011qg,Bastero-Gil:2013nja} (see also \cite{Woodard:2014jba}). More recent works have also tackled this problem from different perspectives: in \cite{Markkanen:2017rvi} it was shown that different renormalizations can be found by expressing the subtraction terms in de Sitter space as counterterms in the Lagrangian; in \cite{Animali:2022lig} the problem of the infrared distortions in adiabatic regularization was dealt by introducing a comoving infrared cutoff; and an alternative method based on a resummation of the entire adiabatic expansion has been proposed in \cite{Corba:2022ugu}.

However, it is crucial to observe that, according to Eq.~(\ref{difreg}), there are infinite equally valid renormalization prescriptions, characterized by different choices of $\alpha$ and $\beta$, that differ from the adiabatic one by geometric contributions. In this work, we use this freedom to propose a new set of subtraction terms that are capable of successfully removing the ultraviolet divergences of the two-point function, while simultaneously minimizing the distortions introduced in the infrared part of the spectrum. More specifically, we wish to construct a regularized power spectrum such that the associated correlation function obeys the following conditions: \vspace{-0.1cm}

\begin{itemize}
    \item[(i)] is regular at the coincident limit $x \rightarrow x'$ , \vspace{-0.2cm}
    \item[(ii)] is equivalent to the general curved-spacetime construction \eqref{renG}, \vspace{-0.2cm}
    \item[(iii)]  recovers $\langleRE \phi^2(x)\rangleRE_{\mathcal{M}}=0$ in Minkowski spacetime,  \vspace{-0.2cm}
\end{itemize}
Moreover, for the regularized power spectrum we ask:\vspace*{-0.1cm}
\begin{itemize}
    \item[(iv)] that it approximates the unregularized one for all amplified infrared modes.
\end{itemize}

Regarding condition (i), the importance of constructing a power spectrum that produces a well behaved correlation function was already stated in \cite{delRio:2014aua}. Indeed, since we wish to encode the information about the quantum vacuum state in the power spectrum, including the consistent construction of $\langle \phi^2(x)\rangle$, it is fundamental to build such a regularized power spectrum. Condition (iii) is introduced to ensure that the standard \textit{normal ordering} prescription is recovered in the flat spacetime limit. Adiabatic regularization satisfies requirements (i)-(iii), but not necessarily (iv). Note that, since adiabatic regularization fulfills condition (ii) \cite{delRio:2014bpa, Beltran-Palau:2021xrk}, any alternative scheme constructed in momentum space that differs from the adiabatic one as \eqref{difreg} will also satisfy (ii).

The rest of this paper is structured as follows. In Sect.~\ref{sec:adiabatic} we present an alternative regularization prescription that fulfills all conditions (i)-(iv) above.  In Sect.~\ref{sec:RenormConds} we discuss our renormalization conditions and fix the extra possible arbitrary parameters. In Sect.~\ref{sec:examples} we apply our formalism to the regularization of the power spectrum of a massive scalar field in de Sitter space, and show that our proposed method tames the distortions in the infrared part of the spectrum. In Sect.~\ref{sec:summary} we summarize and discuss our results. Finally, in Appendix \ref{sec:appA} we examine the possibility of regularizing the power spectrum with a hard infrared cutoff.

\section{Regularization of the two-point function} \label{sec:adiabatic}

The equation of motion of the scalar field $\phi$ in a FLRW spacetime is
\be \phi '' + 2 \frac{a'}{a} \phi' - \nabla^2 \phi + m^2 \phi + \xi R \phi= 0 \ , \label{eq:scalar-eom} \ee
where we have  included a non-minimal coupling $\xi R \phi^2$ to the Ricci scalar $R=6 a'' / a^3$, with $\xi$ the corresponding coupling strength. We can quantize the field by using the standard canonical \cite{Parker:2009uva} or the algebraic quantization approach \cite{Wald:1995yp,Brunetti:2015vmh}. The two point correlation function (\ref{Gcosmo}) at the coincident point $\vec{x} = \vec{x}'$ can be written as
\be \langle \phi^2(\tau) \rangle = \int_0^{\infty} d \, \text{log}\,k \, \Delta_{\phi} (k,\tau) \ , \label{tpf} \ee
where $\Delta_{\phi} (k,\tau)$ is the (rescaled) power spectrum, given by the field modes $\chi_k$ as
\be  \Delta_{\phi} (k,\tau) := \frac{k^3}{2 \pi^2} \mathcal{P}_k (\tau)  \ , \hspace{0.3cm} \mathcal{P}_k (\tau) = \frac{|\chi_k|^2}{a^2 (\tau)} \ .  \ee
Each mode $\chi_k$ satisfies the equation
\bea
\chi_k''+\left(k^2+m^2a^2+\left(\xi-\frac16\right)a^2 R\right)\chi_k=0 \ ,  \label{eq:fieldmodeeq}
\eea
as well as the normalization condition $\chi_k\chi_k^{*'}-\chi_k'\chi_k^*=ia^{-2}$. The integral in \eqref{tpf} diverges at coincident points $\vec{x} \rightarrow \vec{x}'$, so we need to regularize the expression. Following the subtraction in Eq.~(\ref{eq:integral}), we define the regularized power spectrum as
\be
\Delta_{\phi}^{\rm (reg)} (k, \tau) : = \frac{k^3 }{2 \pi^2} (\mathcal{P}_{k} (\tau) - \mathcal{Q}_{k} (\tau) ) \ ,  \label{tpf2}
\ee
and $\langleRE \phi^2(\vec{x},\tau) \rangleRE =\int_0^{\infty} d \log{k} \, \Delta_{\phi}^{\rm (reg)} (k, \tau)$ is the associated regularized two-point function.\newline

{\bf a) Adiabatic regularization:} Adiabatic regularization is based on an adiabatic expansion of the field modes, which allows to identify and subtract the UV-divergent terms from the quantity to be regularized \cite{Parker:1968mv,Parker:1969au,Parker:1971pt}. The expansion is based on the following WKB ansatz for the modes,
\bea
\chi_k \sim \frac{1}{\sqrt{2W^{(n)}_k(\tau)}}e^{-i \int^\tau W_k^{(n)}(\tau')d\tau'}.\label{eq:wkbansatz}
\eea
where $W_k^{(n)}(\tau)=w^{(0)}+w^{(1)}+...w^{(n)}$ is expanded such that the term $w^{(n)}$ is  of $n$th adiabatic order, i.e.~contains $n$ time-derivatives of the scale factor. The zeroth order contribution corresponds to the Minkowski vacuum state
\bea
W_k^{(0)}=\sqrt{k^2+m^2a^2}:=\omega_k.
\eea
Higher orders of $W_k^{(n)}$ can be obtained by substituting ansatz \eqref{eq:wkbansatz} into \eqref{eq:fieldmodeeq} and solving order by order. Given a magnitude constructed from the field modes, we know by dimensional reasoning up to which order we need to expand adiabatically to correctly isolate the UV divergences. The two-point function only requires subtraction up to second adiabatic order, which gives
\begin{align}
\mathcal{Q}_k &:=\left(\frac{1}{2a^2W_k^{(2)}}\right)^{(2)} \label{Hk1} \\
&= \frac{1}{2a^2\omega_k}-\frac{\left(\xi-\frac16\right) R}{4\omega_k^3}-\frac{3}{16}\frac{\omega_k'^2}{a^2\omega_k^5}+\frac{\omega_k''}{8a^2\omega_k^4} \nonumber \ .
\end{align}
The first term completely subtracts the divergence in Minkowski spacetime, while the second one removes the remaining logarithmic divergence in a curved background. The third and fourth terms only contribute to the two-point function with a finite term proportional to $R$.

Second and higher order terms in the adiabatic expansion behave, for momenta $k \gtrsim a m$, as $\sim k^2 \times \omega_k^{-n/2} \sim k^{2-n}$ with $n\geq 3$. Therefore, if the mass of the field is smaller than the maximum comoving momenta amplified by the non-adiabatic expansion of the universe $k_+/a$  (i.e.~$m a \lesssim k_+$), these terms can significantly distort the power spectrum at scales of physical interest. The lighter the field is, the worse the distortions.\newline \vspace*{-0.1cm} 

{\bf b) Regularization without infrared distortions:} However, \eqref{Hk1} is not the only possible set of subtraction terms that can be obtained from an adiabatic expansion. In Ref.~\cite{Ferreiro:2018oxx}, an `off-shell' type of prescription for adiabatic regularization was proposed, in which the zeroth order term of the expansion is changed to
\be
W_k^{(0)}=\sqrt{k^2+\mu^2a^2}:=\tilde{\omega}_k \ , 
\ee
where $\mu$ is now an arbitrary mass scale. By rewriting the field mode equation as
\be
\chi_k''+\left(\tilde{\omega}_k^2-\mu^2a^2+m^2a^2+\left(\xi-\frac16\right)a^2 R\right)\chi_k=0 , \label{eq:fieldmodeeq2}
\ee
we can obtain higher order terms of the expansion by substituting \eqref{eq:wkbansatz} into \eqref{eq:fieldmodeeq2} and solving order by order. In the new prescription, $\tilde \omega_k$ is taken to be of zeroth order, while $\mu^2$ and $m^2$ are of second order (i.e.~the same order as $R$). The subtraction terms for the two-point function obtained with this prescription are now
\be
\widetilde{Q_k}:=\frac{1}{2a^2\tilde{\omega}_k}+\frac{(\mu^2-m^2)}{4\tilde{\omega}_k^3}-\frac{(\xi-\frac16)R }{4\tilde{\omega}_k^3}-\frac{3}{16}\frac{\tilde{\omega}_k'^2}{a^2 \tilde{\omega}_k^5}+\frac{\tilde{\omega}_k''}{8a^2\tilde{\omega}_k^4}\label{Hk2} \ ,
\ee
and coincide with the ones of the traditional adiabatic approach \eqref{Hk1} only for $\mu = m$. The difference between the regularized two-point functions obtained with the traditional and `off-shell' adiabatic prescriptions (which we denote as $\langleRE \phi^2(x)\rangleRE$ and $\langleRE \widetilde{\phi^2(x)} \rangleRE$ respectively) is
\bea
&&\langleRE \widetilde{\phi^2(x)}\rangleRE-\langleRE \phi^2(x) \rangleRE = \\
&&\frac{1}{16\pi^2}\left [ \mu^2-m^2 \left(m^2+\left(\xi-\frac16\right)R\right) \log{\left(\frac{m^2}{\mu^2}\right)} \right] \nonumber  \ .
\eea
The difference is given by geometric terms as in Eq.~\eqref{difreg}, so the `off-shell' prescription also satisfies property (ii).

By setting $\mu$ large enough, we can tame the infrared distortions introduced by subtraction terms \eqref{Hk2} in the power spectrum, while still removing the ultraviolet divergent part. However, this prescription does not yield $\langleRE \phi(x) \rangleRE_{\mathcal{M}}=0$ for $\mu \neq m$ in Minkowski spacetime (i.e.~does not obey condition iii above). We can solve this problem by imposing $\mu = m$ only in the Minkowskian part of the subtraction [i.e.~the first two terms in \eqref{Hk2}], while still retaining the freedom of choosing $\mu$ in the other terms. Moreover, the last two terms in \eqref{Hk2} yield a finite contribution proportional to $R$ to the regularized two point function, so we can remove them while still being compatible with Eq.~\eqref{difreg}. We then propose the following subtraction terms,
\bea
\overline{\mathcal{Q}_k}:=\frac{1}{2a^2\sqrt{k^2+m^2a^2}}-\frac{(\xi-1/6)R}{4(k^2+M^2a^2)^{3/2}}\label{Hk3}
\eea
where $M$ is an arbitrary mass scale that plays the same role as $\mu$ in \eqref{Hk2}. The difference between the regularized two-point functions obtained  with the traditional adiabatic prescription and the one obtained by subtracting \eqref{Hk3} (which we denote as $\langle {:}\,\overline{\phi^2(x)}\,{:}\rangle$) is
\be
\langle {:}\,\overline{\phi^2(x)}\,{:}\rangle-\langle{:}\,\phi^2(x)\,{:}\rangle=\frac{\xi-\frac16}{16\pi^2} \log{\left(\frac{m^2}{M^2}\right)}R+\frac{R}{288\pi^2}\label{diff2}
\ee
which again can be written as a sum of geometrical terms as in \eqref{renG}. Therefore, our proposed regularization prescription also satisfies condition (ii). Note that the last term in \eqref{diff2} is a consequence of having removed the last two subtraction terms in \eqref{Hk2}. 

In summary, let us define the regularized two point function for a given choice of $M$ as
\be \langleRE \overline{ \phi^2} \rangleRE_M := \int_0^{\infty} d \, \text{log}\,k \, \frac{k^3 }{2 \pi^2} (\mathcal{P}_{k}  - \mathcal{\overline{Q}}_{k}  ) \ . \label{tpf3} \ee

\section{Renormalization conditions and coupling constants } \label{sec:RenormConds} 

In order to give physical meaning to our regularized two point function \eqref{tpf3}, we need to fix the coupling constants of our theory at a given mass scale. In our case, the gravitational couplings have an explicit appearance in the dynamical equations of the scale factor
\bea
\frac{1}{8\pi G}R-4\Lambda-\alpha \Box R= \langleRE T^a_a\rangleRE \ ,  \label{semic}\\
\nabla_a\langleRE T^{ab}\rangleRE=0 \ , 
\eea
where $\langleRE T^{ab} \rangleRE $ is the regularized stress-energy tensor and  $\{G, \Lambda, \alpha \}$ are coupling constants. Here we have added the higher curvature-order term $\alpha \Box R$ in order to fully reabsorb the proper divergence of the stress-energy tensor. The conservation of the stress-energy tensor is  automatically implied by construction. The trace of the regularized stress energy tensor can be written as \cite{Brunetti:2015vmh}
\begin{align}
\langleRE T_a^a \rangleRE =&\left(3\left(\xi-\frac16\right)\Box+m^2\right)\langleRE \overline{ \phi^2}\rangleRE_M  +T_{\rm an}\nonumber \\&+a+bR+c\Box R\ , \label{Tren}
\end{align}
where $T_{\rm an}$ is the usual contribution to the trace anomaly given by
\be
T_{\rm an} \equiv \frac{1}{2880\pi^2}\left(R_{ab}R^{ab}-\frac13 R^2\right)-\frac{1}{32\pi^2}\left(\xi-\frac16\right)^2R^2 \ , \label{anom}
\ee
and $\{a, b, c\}$ are three arbitrary constants that parameterize the ambiguities of the regularized stress-energy tensor \cite{Wald:1995yp}. Note that this contribution \eqref{anom} cannot be re-absorbed in any of these ambiguities.

Both $m$ and $\xi$ are not renormalized in a free theory, so they take the physical values. Therefore, we just need to fix the gravitational parameters $\{\Lambda, G , \alpha\}$ on the left hand side of Eq.~\eqref{semic}, as well as $\{a, b, c\}$ and $M$ on the right hand side.

Let us impose the conditions at our current universe, where both $G$ and $\Lambda$ are measured. We consider first the case $M=m$. Assuming $m \gg R$ today, we can use the adiabatic expansion of the exact modes to approximate the two point function as,
\be
 \langleRE \overline{ \phi^2}\rangleRE_m  =-\frac{R}{288\pi^2}+\frac{(\xi-\frac16)}{32\pi^2m^2}R^2-\frac{(\xi-\frac16)}{16\pi^2m^2}\Box R+\mathcal{O}(m^{-4})\label{phiadb2}
\ee
By substituting \eqref{phiadb2} into \eqref{Tren}, and this one into \eqref{semic}, we can see that a convenient prescription is $b=-\frac{m^2}{288\pi^2}$ $a=0$ and $c=\frac{5(\xi-1/6)}{96\pi^2}$. In this case we have 
\bea
\frac{1}{8\pi G_{\rm o}}R+4\Lambda_{\rm o}+\alpha_{\rm o} \Box R=  \langleRE \roverline{T^a_a} \rangleRE_m   \ , \,\,\,\,\,\,\,\,  \\
  \langleRE  \roverline{T_a^a} \rangleRE_m  : = \frac{1}{2880\pi^2}\left(R_{ab}R^{ab}-\frac13 R^2\right)+\mathcal{O}\left(m^{-2}\right) \nonumber \ ,
\eea
where $\{ \Lambda_{\rm o} , G_{\rm o} ,\alpha_{\rm o} \}$ are the observed physical values of the gravitational coupling constants since the extra contributions coming from the stress-energy tensor are negligible.

At any other instant, the stress energy tensor is then fixed as
\bea
 \langleRE  \roverline{T_a^a} \rangleRE_m=&&\left(3\left(\xi-\frac16\right)\Box+m^2\right)\langleRE \overline{ \phi^2}\rangleRE_m  +T_{\rm an}\nonumber \\&&-\frac{m^2}{288\pi^2}R+\frac{5(\xi-1/6)}{96\pi^2}\Box R. 
\eea
 If we now choose a different prescription $M \neq m$, the trace of the regularized stress-energy tensor changes as
\begin{align}
 \langleRE  \roverline{T_a^a} \rangleRE_M =& \langleRE  \roverline{T_a^a} \rangleRE_m +\frac{3(\xi-1/6)}{16\pi^2}\log{\left(\frac{m^2}{M^2}\right)}\Box R\nonumber \\&+\frac{\xi-1/6}{16\pi^2}\log{\left(\frac{m^2}{M^2}\right)}R \ ,
\end{align}
and in order to preserve the form of Eq.~(\ref{semic}), the gravitational constants must change as
\begin{align}
\alpha_M &=\alpha_{\rm o}+\frac{3(\xi-1/6)}{16\pi^2}\log{\left(\frac{m^2}{M^2}\right)} \ , \\
G_M &=\frac{G_{\rm o}}{1-\frac{G_{\rm o}(\xi-1/6)m^2}{2\pi}\log{\left(\frac{m^2}{M^2}\right)}} \ ,
\end{align}
whereas $\Lambda_0$ does not receive any contribution. Note that the change in $G$ is negligible for a field with $m \ll m_{\rm Pl}$. Taking into account that we can safely neglect the possible higher order correction of the curvature (except at very early times), we can assume that a possible rescaling of the regularized two point function will not alter the gravitational coupling sufficiently to have a effective impact.

\section{Regularized power spectrum in de Sitter spacetime} \label{sec:examples}

\begin{figure*}
    \centering
    \includegraphics[width=0.49\textwidth]{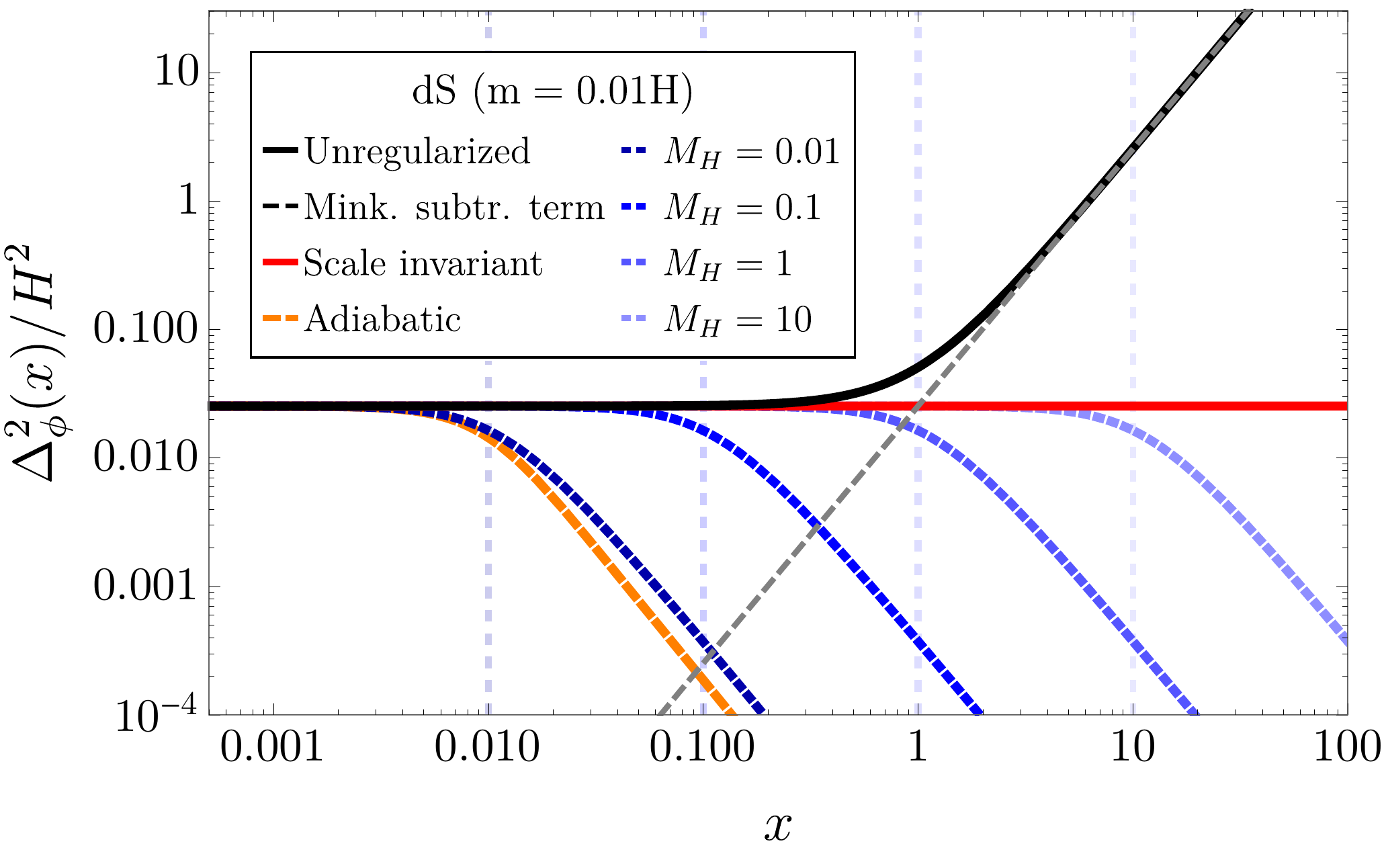} \,\,
    \includegraphics[width=0.49\textwidth]{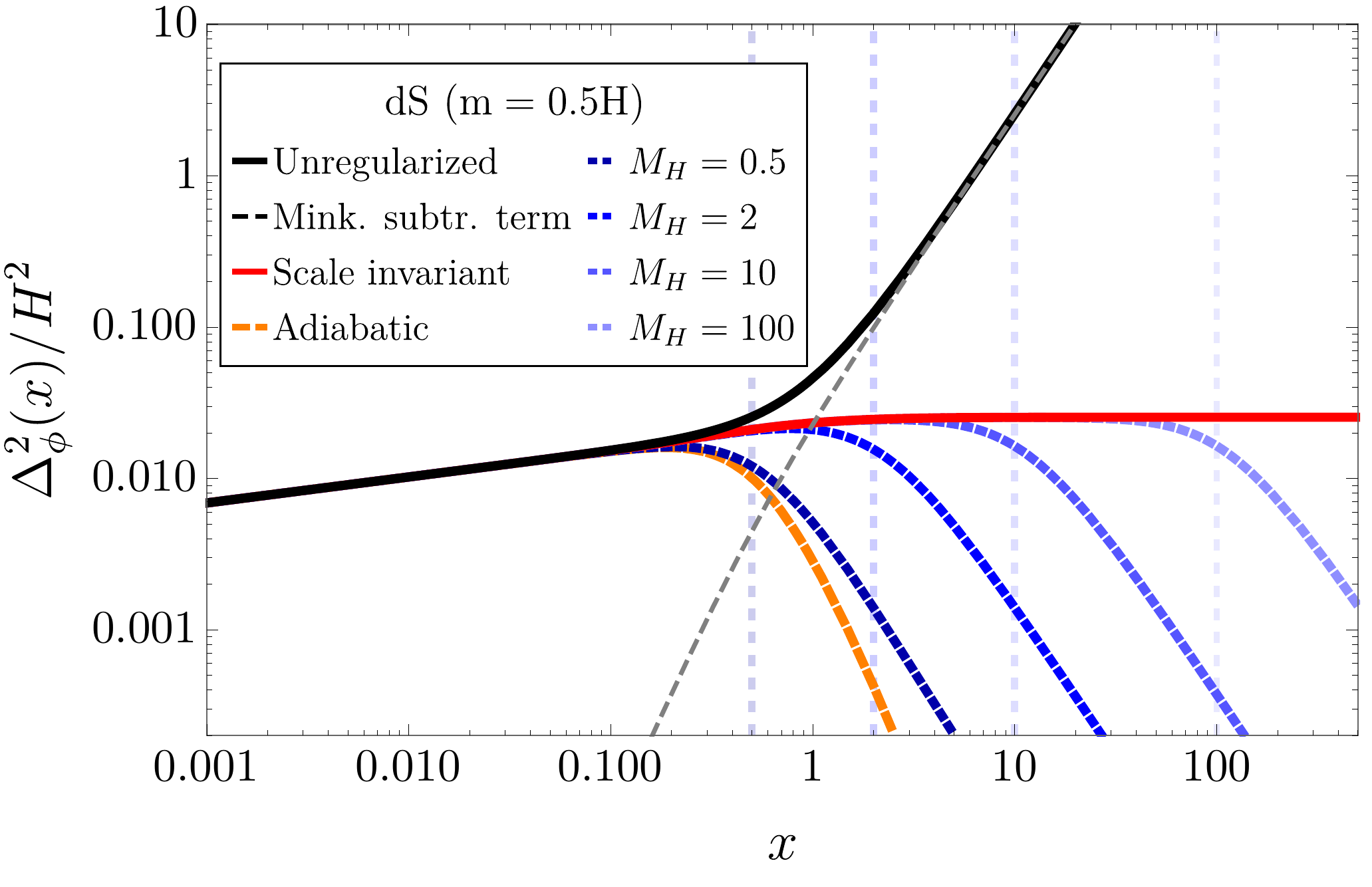}
    \caption{Power spectrum of a scalar field in de Sitter space with $m = 0.01 H$ (left) and $m = 0.5H$ (right). We show the unregularized spectra (\ref{eq:unreg-pws}) (black), the Minkowskian subtraction term $\propto (m_H^2 + x^2)^{-1/2}$ (grey dashed), the quasi scale-invariant spectrum obtained by subtracting the Minkowskian term from (\ref{eq:unreg-pws}) (red), the spectra regularized with the traditional adiabatic approach (\ref{eq:powsp-dS-trad}) (orange dashed), and the one regularized with our proposed subtraction terms \eqref{Hk3} for different values of $M_H$ (blue dotted). The values of $M_H$ are also indicated with dashed vertical bars.} \vspace*{-0.2cm} \label{fig:example-desitter}
\end{figure*}

In order to illustrate the differences between regularization prescriptions, let us consider a massive scalar field with $\xi= 0$ in de Sitter spacetime.  The scale factor evolves as $a(\tau) = - (H \tau)^{-1}$ with (constant) Hubble parameter $H$, and the field mode equation (\ref{eq:fieldmodeeq}) becomes
\be \chi_k''  + \left( k^2 + \frac{m_H^2 - 2 }{\tau^2} \right) \chi_k = 0 \ , \hspace{0.4cm} m_H \equiv \frac{m}{H} \label{equbd}\ . \ee
A natural solution is the Bunch-Davies vacuum \cite{Bunch:1978yq}\footnote{If $m > (3/2)H$, the solution must include an extra factor $e^{-\frac{\pi}{2} Im[\nu]}$ in order to be correctly normalized.}
\be \chi_k = \frac{-i\sqrt{\pi \tau}}{2}  H_{\nu}^{(1)} (-k \tau) \ , \hspace{0.3cm} \nu \equiv \sqrt{\frac{9}{4} - m_H^2}  \ , \label{eq:fieldsol} \ee
and the unregularized power spectrum is in this case
\be \Delta_{\phi} (k, \tau) = \frac{H^2 x^3}{8 \pi} |H_{\nu}^{(1)} (x) |^2  \ , \hspace{0.3cm} x \equiv \frac{k}{H a} \ , \label{eq:unreg-pws} \ee
where $H_{\nu}^{(1)}$ is the Hankel function of the first kind. In the $m=0$ case we simply have $\Delta_{\phi} (x) = H^2 (1 + x^2 ) /(4 \pi^2)$.

Adiabatic regularization yields the following regularized power spectrum [$ \Delta_{\phi}^{\rm (reg)} \equiv \Delta_{\phi} - k^3 \mathcal{Q}_{k}/(2 \pi^2 )$] \cite{Parker:2007ni}
\begin{align}
\Delta_{\phi}^{\rm (reg)}& =  \frac{H^2 x^3}{8 \pi} \left( |H_{\nu}^{(1)} (x) |^2 - \frac{g (x; m_H)}{4 \pi  \left(m_H^2+x^2\right)^{7/2}} \right) , \nonumber \\[5pt] 
 & g (x; m_H) \equiv 8 x^6 +8 x^4 \left(3 m_H^2 + 1\right)+\label{eq:powsp-dS-trad}\\
 & \hspace{0.9cm} 2 m_H^2 x^2 \left(11+12 m_H^2\right)+m_H^4 \left(9+8 m_H^2\right) , \nonumber 
\end{align}
which gives exactly $\Delta_{\phi}^{\rm (reg)} = 0$ for $m=0$. Regularizing the two-point function with our proposed subtraction terms (\ref{Hk3}) yields instead the following result [$\overline{\Delta}_{\phi}^{\rm (reg)} \equiv \Delta_{\phi} - k^3 \overline{\mathcal{Q}}_{k}/(2 \pi^2 )$], 
\be \overline{\Delta}_{\phi}^{\rm (reg)} = \frac{H^2 x^3}{8 \pi} \left( |H_{\nu}^{(1)} (x) |^2 -  \frac{2\pi^{-1}}{ \sqrt{m_H^2 + x^2}}  - \frac{2\pi^{-1}}{(M_H^2 + x^2)^{\frac{3}{2}}}  \right)  \label{eq:powsp-dS-alt} \ee
with $M_H \equiv M/H$. The power spectrum in the massless case $m=0$ is now
\be
\overline{\Delta}_\phi^{\rm (reg)} =\frac{H^2}{4\pi^2}\left[1+\left(1+\frac{M_H^2}{x^2}\right)^{-3/2}\right] \ ,\ee
which recovers the scale-invariant result $\overline{\Delta}_\phi^{\rm (reg)} \simeq H^2 /(4 \pi^2)$ at infrared scales\footnote{Note that the Bunch-Davies vacuum generates a well-known infrared divergence in the two-point function, see e.g.~\cite{Ford:1977in}.} $x \ll M_H$.

In the left panel of Fig.~\ref{fig:example-desitter} we compare the power spectra obtained with different regularization prescriptions, for the choice $m=0.01 H$. In red we depict the quasi scale-invariant power spectrum obtained by subtracting
only the first term in (\ref{Hk3}) to the unregularized expression (\ref{eq:unreg-pws}), as we would do in Minkowski spacetime. The regularized two-point function associated to such spectra is UV divergent. In orange we show the regularized power spectrum obtained with the traditional adiabatic approach (\ref{eq:powsp-dS-trad}): such prescription removes the UV divergences, but significantly suppresses the spectrum at scales $m_H \lesssim x (\lesssim 1)$. The different blue lines show the power spectra regularized with our proposed subtraction terms \ref{eq:powsp-dS-alt} for different choices of $M_H \equiv M /H$. By setting $M_H = 1$, we can recover the quasi scale-invariant power spectrum at all superhorizon modes $x \lesssim 1$ while removing the UV tail at $x \gtrsim 1$. The regularized two-point functions obtained from the integration of these spectra are finite. The right panel of Fig.~\ref{fig:example-desitter} depicts the same comparison for the heavier field $m = 0.5 H$, which illustrates that all regularized spectra recover the characteristic tilt at scales $x \ll 1$ independently of the choice of $M_H$.

\section{Summary and conclusions} \label{sec:summary}

In this work we have proposed an alternative regularization prescription for the two-point function of a scalar field in a FLRW spacetime, consisting in subtracting (\ref{Hk3}) from the integrand of the two-point function in momentum space, see Eq.~(\ref{eq:integral}). Unlike adiabatic regularization, our proposed subtraction terms are designed to subtract the UV-divergent tail of the power spectrum while minimizing the distortions at infrared scales, meaning the part of the spectrum amplified by the non-adiabatic expansion of the universe. Our prescription depends on a free mass scale $M$ that acts as a soft cutoff, and differs from traditional adiabatic regularization as in Eq.~(\ref{difreg}). As a consequence it is constructed in a local and covariant way. We have illustrated our method by regularizing the power spectrum of a scalar field in de Sitter space: by setting $M \approx H$, we can recover the standard result $\Delta_{\phi}^{\rm (reg)} \simeq H^2 /(4\pi^2)$ for all superhorizon modes, while simultaneously removing the UV tail for subhorizon ones. 

In future work we plan to reexamine the regularization of the stress-energy tensor, which in curved spacetime is also not unique. Similarly to the two-point function, the regularized stress-energy tensor is not uniquely defined, so we could potentially use the intrinsic ambiguities of the renormalization program in order to find a set of subtraction terms that also minimize the infrared distortions. Our proposed regularization prescription can potentially also be extended to other scenarios such as fields with interactions to homogeneous time-dependent fields, or to other field species like fermions or gauge fields. We plan to study these topics elsewhere.\vspace*{-0.2cm}

\,\newline
{\bf Acknowledgements:} We thank Jose Navarro-Salas for very useful remarks about the manuscript. A. F. is supported by the Irish Research Council Postdoctoral Fellowship No.~GOIPD/2021/544. F.T. is supported by a \textit{Mar\'ia Zambrano} grant (ZA21-034) from the Spanish Ministry of Universities and grant PID2020‐116567GB‐C21 of the Spanish Ministry of Science. \vspace*{-0.3cm}

\appendix
\section{On the regularization of the two-point function with a hard cutoff} \label{sec:appA}

The subtraction terms \eqref{Hk3} proposed in this work for the regularized two-point function act as a soft cutoff: they suppress the power spectrum only for momenta $k \gtrsim M$, as seen in Fig.~\ref{fig:example-desitter} for de Sitter spacetime. A natural question to ask is whether we can regularize with a \textit{hard} cutoff instead. Let us then consider the possibility of subtracting the terms \eqref{Hk1} obtained with the traditional adiabatic prescription, but only for momenta $k > k_-$ with $k_-$ a hard comoving infrared cutoff. Let us denote the two-point function regularized this way as $\overline{\langleRE \phi^2\rangleRE }$, and compare it with the one regularized with adiabatic regularization, in which all momenta down to $k=0$ have been subtracted. The difference is given by
\begin{align} \allowdisplaybreaks
\langleRE \phi^2 \rangleRE - \overline{\langleRE \phi^2 \rangleRE} & =\frac{4\pi}{a^2(2\pi)^3}\int^{ak _-}_0  \mathcal{Q}_k(\tau,m)\\
&=A(k_-)+B(k_-)H^2(\tau)+C(k_-)R(\tau) \nonumber \end{align}
where $H(\tau)=a'/a^2$, $R(\tau) = 6 a''/a^3$, and \{$A(k_-)$, $B(k_-)$, $C(k_-)$\} are three functions that depend on the infrared cutoff as follows,
\begingroup
\allowdisplaybreaks
  \begin{align}
A(k_-)=&\frac{1}{8\pi^2} \left[k_-\sqrt{k_-^2+m^2} +m^2\log(m) \right.  \\
&\left. \hspace{1cm} - m^2 \log \left( k_-+\sqrt{k_-^2+m^2} \right) \right] \ , \nonumber \\
B(k_-)=&-\frac{k_-^3m^2}{32\pi^2(k_-^2+m^2)^{5/2}} \ ,\\
C(k_-)=&\frac{1}{48\pi^2}\left[\frac{6k_-m^2(6\xi-1)+k_-^3(36\xi-5)}{(k_-^2+m^2)^{3/2}} \right.  \label{eq:Ckminus} \\
&\left. \hspace{-1cm} +6(6\xi-1) \left(\log (m) - \log 
 \left( k_-+\sqrt{k_-^2+m^2}\right) \right) \right] \nonumber  \ .\end{align} 
 \endgroup
We must have $B(k_-) = 0$ in order to satisfy Eq.~\eqref{difreg}, which for a massive field only happens for exactly $k_{-}=0$. In consequence, regularization with a hard cutoff is not in general covariant. For $m=0$, adiabatic regularization introduces an infrared divergence (see the behaviour of Eq.~\eqref{eq:Ckminus} in the limit $m \rightarrow 0$), so one needs to add an additional squared mass scale $\mu^2 > 0$ into the scheme and the argument is analogue. \newpage

  \bibliography{References.bib}
  \bibliographystyle{apsrev4-1}

\end{document}